\documentclass[9.5pt]{article}
\usepackage{array, amsmath, amssymb}
\usepackage{multicol}
\begin{document}
\begin{center}
{\Large\bf Epidemic thresholds in directed complex networks}\\
\vspace{0.5cm}
{\large\bf Shinji Tanimoto}\\
{\texttt{tanimoto@cc.kochi-wu.ac.jp}}\\
\vspace{0.5cm}
Department of Mathematics,
Kochi Joshi University,
Kochi 780-8515, 
Japan. \\
\end{center}
\begin{abstract}
The spread of a disease, a computer virus or information is discussed in a directed complex network. 
We are concerned with a steady state of the spread for the SIR and SIS dynamic models.
In a scale-free directed network it is shown that the threshold of its outbreak in both models
approaches zero under a high correlation between nodal indegrees and outdegrees.
\end{abstract}
\vspace{0.3cm}
\begin{multicols}{2}
\begin{center}
{\bf\large 1. Introduction}  \\
\end{center}
\indent
\indent
Recent studies on the spread of diseases, computer viruses or information
in complex networks have been exclusively
devoted to undirected ones such as social networks or the Internet. 
In those networks the direction of links (or edges) 
is not important and can be ignored. On the other hand, there are important directed networks
in nature and man-made systems such as food webs and the WWW, {\it etc}. However, studies on the
spread 
in directed networks have not been done extensively. In order to tackle the problem, [5] made good use of 
the generating function methodology. Our approach is different from it 
and is based on the dynamical mean-field theory of [1, 2, 6]. \\ 
\indent
The objective is to study the SIR and SIS models in directed networks and to
derive the critical infection rate or the threshold, above which a disease spreads in the networks and
below which it dies out. 
In a directed network a disease passes to other nodes through
outgoing links and a node is infected by incoming links. The indegree of a node is the number of
incoming links into the node and the outdegree is that of
outgoing links emanating from it. \\
\indent
Many directed networks such as the WWW, networks of metabolic reactions and phone calls 
have power law degree distributions ([4], [8], [10]): 
\[
P(k) \propto k^{-\gamma}~~ {\rm and} ~~Q(\ell) \propto {\ell}^{-\gamma'}
\]
for all large indegrees $k$ and outdegrees $\ell$, respectively, 
although real networks inevitably have finite sizes
of $k, \ell$. The distinction between both degree distributions disappears for an undirected network.
If the exponents satisfy $2 < \gamma, \gamma' \le 3$,
these networks are called scale-free. 
Several authors use $\gamma_{\rm in}, \gamma_{\rm out}$
in place of $\gamma, \gamma' $, respectively.\\
\indent
In [9] it was shown that the threshold of the SIS epidemic model in undirected scale-free networks 
is zero.
In [1, 2, 6] a similar result was obtained also for the SIR epidemic model. \\
\indent
In this paper, using the joint probability distribution of indegrees and outdegrees, 
we derive the thresholds for the SIR and SIS epidemic dynamics 
on directed networks. Actually they turn out the same for both. Furthermore, it is shown that 
the threshold approaches zero  
under a high correlation between indegrees and outdegrees.
In the SIR model, the average fraction of nodes that are ever infected
until the disease dies out is also given using the indegree distribution. \\
\begin{center}
{\bf\large 2. The SIR model in a directed \\
network} \\
\end{center}
\indent
\indent
First we investigate the SIR model on a directed network. 
Nodes of the network are divided into the following three groups as in [7, Chap. 10]: 
Susceptible (S), Infected (I) and Removed (R). Hereafter we will denote a susceptible node by
an S-node {\it etc.,} for short.
An S-node becomes infected at a rate of $\lambda$ ($0 \le \lambda \le 1$). 
The parameter $\lambda$ is the infection rate, for which we will derive the critical value 
for an outbreak of a disease, a computer virus, {\it etc.}
The disease can be passed from I-nodes to S-nodes following only the direction of directed links. 
R-nodes have either recovered from the disease or died and so they cannot pass the disease to other nodes.
An I-node becomes an R-node at a rate $\delta$ ($0 \le \delta \le 1$) and, without loss of generality, 
we will set $\delta = 1$. \\
\indent
Let us denote the densities of S-, I-, R-nodes with indegree $k$ and outdegree $\ell$ at time $t$ 
by $S_{k,\ell}(t), \rho_{k,\ell}(t), R_{k,\ell}(t)$, respectively. 
So we have 
\[
S_{k,\ell}(t) + \rho_{k,\ell}(t) + R_{k,\ell}(t) =1.
\]
\indent
Let $p(k, \ell)$ be the joint probability distribution of nodes with indegree $k$ and outdegree $\ell$,
and let us denote the marginal distributions by
\[
P(k) = \sum_{\ell} p(k, \ell), ~~ Q(\ell) = \sum_{k} p(k, \ell)
\]
and the averages including the $n$th moments by
\begin{eqnarray*} 
\langle k^n\rangle  \! \! \! \!& = & \! \!\! \!  \sum_{k, \ell} k^n p(k, \ell) = \sum_{k} k^n P(k),  \\
\langle{\ell}^n\rangle  \! \! \! \!& = & \! \!\! \! \sum_{k, \ell} {\ell}^n p(k, \ell) = \sum_{\ell} {\ell}^n Q(\ell), \\
\langle{k\ell}\rangle \! \! \! \!& = & \! \!\! \! \sum_{k, \ell} k\ell p(k, \ell).
\end{eqnarray*} 
\indent
Following the dynamical mean-field theory ([2, 6]), we see that
the spreading process on a directed network can be described by the system of differential equations:
\begin{eqnarray} 
\frac{dS_{k,\ell}}{dt} \! \! \! \!& = & \! \!\! \! - \lambda k S_{k,\ell}(t)\theta(t), \\
\frac{d{\rho}_{k,\ell}}{dt} \! \!\! \! & =  &\! \! \! \! \lambda k S_{k,\ell}(t)\theta(t) - {\rho}_{k,\ell}(t), \\
\frac{dR_{k,\ell}}{dt} \! \!\! \!  & = &\! \!\! \!  {\rho}_{k,\ell}(t).
\end{eqnarray} 
The term $\lambda k S_{k,\ell}(t)\theta(t)$
in (1) and (2) indicates the fraction of newly infected nodes through $k$ incoming links. 
The probability, $\theta(t)$, that a randomly selected outgoing link emanates from an I-node 
at time $t$ is given by
\begin{eqnarray} 
\theta(t) = \frac{\displaystyle \sum_{k, \ell}\ell  
p(k, \ell){\rho}_{k,\ell}(t)}{\displaystyle \sum_{k, \ell}{\ell}p(k, {\ell})}=
\frac{ \sum_{k, \ell} \ell p(k, \ell) {\rho}_{k,\ell}(t)}{\langle{\ell}\rangle}.
\end{eqnarray} 
Note that each directed link is counted twice as one 
outdegree of some node and as one indegree of another. Hence the average outdegree is equal to
the average indegree: ${\langle{\ell}\rangle} = {\langle{k}\rangle}$. \\
\indent
Using the initial condition $S_{k,\ell}(0) = 1$, (1) is easily solved as 
\begin{eqnarray}
S_{k,\ell}(t) = e^{-\lambda k \phi(t)}, 
\end{eqnarray}
where 
\begin{eqnarray*}
\phi(t) = \int_0^t \theta(t')dt'.
\end{eqnarray*}
By (4) and $R_{k, \ell}(0) = 0$, $\phi(t)$ is expressed as
\begin{eqnarray}
\phi(t)  \! \! \! \!& = 
&\! \! \! \!  \frac{1}{{\langle{\ell}\rangle}}
 \sum_{k, \ell} \ell p(k, \ell) \int_0^t {\rho}_{k,\ell}(t')dt' \nonumber \\
\! \! \! \!& = 
&\! \! \! \!  \frac{1}{{\langle{\ell}\rangle}} \sum_{k, \ell} \ell p(k, \ell) R_{k,\ell}(t).
\end{eqnarray}
\indent
We derive the differential equation for $\phi(t)$. Using (3), (5) and (6), it follows that
\begin{eqnarray*}
\frac{d\phi(t)}{dt} \! \! \! \!& = 
&\! \! \! \! \frac{1}{{\langle{\ell}\rangle}} \sum_{k, \ell} \ell p(k, \ell) \frac {dR_{k,\ell}(t)}{dt} \\
\! \! \! \!& = 
&\! \! \! \! \frac{1}{{\langle{\ell}\rangle}} \sum_{k, \ell} \ell p(k, \ell) {\rho_{k,\ell}(t)}\\
\! \! \! \! & = 
&\! \! \! \! \frac{1}{{\langle{\ell}\rangle}} \sum_{k, \ell} \ell p(k, \ell) (1- R_{k,\ell}(t) - S_{k,\ell}(t) ) \\
\! \! \! \! & = 
&\! \! \! \!  1 -\phi(t) - \frac{1}{{\langle{\ell}\rangle}} \sum_{k, \ell} \ell p(k, \ell) S_{k,\ell}(t) \\
\! \! \! \!& = 
&\! \! \! \! 1- \phi(t) - \frac{1}{{\langle{\ell}\rangle}} \sum_{k, \ell} \ell p(k, \ell) e^{-\lambda k \phi(t)}. 
\end{eqnarray*}
\indent
In this paper we are concerned with a steady state of the epidemic spreading.
At the steady state we will have a limit 
\begin{eqnarray*}
\Phi = \lim_{t \rightarrow \infty}\phi(t),
\end{eqnarray*}
together with the condition
\begin{eqnarray*}
 \lim_{t \rightarrow \infty} \frac{d\phi(t)}{dt} = 0.
\end{eqnarray*}
Substituting these into the above equations yields the equation for $\Phi$ as follows:
\begin{eqnarray}
\Phi = 1- \frac{1}{{\langle{\ell}\rangle}} \sum_{k, \ell} \ell p(k, \ell) e^{-\lambda k \Phi}. 
\end{eqnarray}
\indent
An epidemic outbreak implies that this equation has a solution $\Phi >0$ other than $\Phi = 0$. 
Since the right hand side of (7) is a concave function of $\Phi$ and its value at $\Phi = 1$ is less than 1, 
the condition for it is
\[
\frac{d}{d \Phi}\bigg(1- \frac{1}{{\langle{\ell}\rangle}} \sum_{k, \ell} \ell p(k, \ell) 
e^{-\lambda k \Phi}\bigg)_{\Phi = 0} > 1.
\]
Hence the critical infection rate $\lambda_{\rm c}$ or the threshold for an epidemic outbreak is obtained by
setting
\[
\frac{d}{d \Phi}\bigg(1- \frac{1}{{\langle{\ell}\rangle}} \sum_{k, \ell} \ell p(k, \ell) 
e^{-\lambda k \Phi}\bigg)_{\Phi = 0} = 1.
\]
Solving this for $\lambda$ we get
\begin{eqnarray}
{\lambda}_{\rm c} = \frac{\langle{\ell}\rangle}{\sum_{k, \ell} k\ell p(k, \ell)} 
= \frac{\langle{\ell}\rangle}{\langle{k \ell}\rangle}.
\end{eqnarray}
\\
\indent
In [7, Chap.10] the total number of infected individuals is discussed for the classical SIR model,
which represents the final outbreak size.
In our setting it is the average fraction of nodes ever infected
until the disease dies out. This is written as
\begin{eqnarray}
{\mathcal{O}}  \! \! \! \!& = &\! \! \! \! \sum_{k, \ell} p(k, \ell)(1- S_{k, \ell}(\infty)) \nonumber\\
\! \! \! \!& =  &\! \! \! \! 1- \sum_{k, \ell} p(k, \ell)S_{k, \ell}(\infty)  \\
\! \! \! \!& =  &\! \! \! \! 1- \sum_{k, \ell} p(k, \ell)e^{-\lambda k \Phi}  \nonumber
=  1- \sum_{k} P(k)e^{-\lambda k \Phi}  
\end{eqnarray}
by means of the indegree distribution $P(k)$.\\
\indent
Suppose that the indegree distribution follows a power law 
\[
P(k) \propto k^{-\gamma}, ~~k \ge m
\] 
with $2<\gamma \le 3$, then $P(k) = 
(\gamma-1)m^{\gamma-1}k^{-\gamma}~(k \ge m)$,
where $m$ is the minimum indegree.
Let $\Gamma(a,x)$ be the incomplete gamma function defined by
\[
\Gamma(a,x)= \int_x^{\infty}t^{a-1}e^{-t}dt.
\]
Then the fraction (9) of the outbreak size can be written as
\begin{eqnarray*}
{\mathcal{O}} \! \! \! \!& = 
&\! \! \! \!  1 - \int_m^{\infty}P(k)e^{-\lambda k \Phi}dk \nonumber \\
\! \! \! \!& = 
&\! \! \! \! 1- (\gamma -1)(\lambda m \Phi)^{\gamma - 1}
\Gamma(1-\gamma, \lambda m \Phi),
\end{eqnarray*}
by the continuous approximation.
\\
\begin{center}
{\large \bf 3. The SIS model in a directed \\
network}  \\
\end{center}
\indent
\indent
In the SIS model R-nodes are absent. Those nodes that recovered from a disease may be
infected again and again.
The densities of S-, I-nodes with indegree $k$ and outdegree $\ell$ at time $t$ are
$S_{k,\ell}(t), \rho_{k,\ell}(t)$ as before, and the equality $S_{k,\ell}(t) = 1- \rho_{k,\ell}(t)$ holds.
So equations (1)-(3) are replaced by the single differential equation
\begin{eqnarray} 
\frac{d{\rho}_{k,\ell}}{dt}  =   \lambda k (1- {\rho}_{k,\ell}(t))\theta(t) - {\rho}_{k,\ell}(t), 
\end{eqnarray} 
where $\theta(t)$ is the same probability as (4). \\
\indent
At the steady state, as in Section 2, we will have the condition
\[
\lim_{t \rightarrow \infty}\frac{d{\rho}_{k,\ell}}{dt}  = 0
\]
for all $k$ and $\ell$, and a limit
\[
\Theta = \lim_{t \rightarrow \infty}\theta(t).
\]
So we get from (10)
\[
  \lim_{t \rightarrow \infty} {\rho}_{k,\ell}(t)= \frac{\lambda k \Theta}{1+ \lambda k \Theta}. 
\]
\indent
Substituting these into (4) we have the equation for $\Theta$ as follows:
\[
\Theta = 
\frac{1}{\langle{\ell}\rangle} \sum_{k, \ell} \ell p(k, \ell)  \frac{\lambda k \Theta} {1+ \lambda k \Theta}.
\]
If this has a solution $\Theta > 0$ other than $\Theta = 0$, then it corresponds to an endemic state.
Since the right hand side of the equation is a concave function of $\Theta$ and its value at $\Theta = 1$ is less than 1,
the condition for an outbreak is
\[
\frac{d}{d \Theta}\bigg(\frac{1}{\langle{\ell}\rangle} \sum_{k, \ell} \ell p(k, \ell)  
\frac{\lambda k \Theta} {1+ \lambda k \Theta}\bigg)_{\Theta = 0} >1.
\]
Again, this yields the same threshold (8) as in the SIR model:
\[
{\lambda}_{\rm c} = \frac {\langle{\ell}\rangle}{\langle{k \ell}\rangle}.
\]
\indent
In the next section we calculate the threshold ${\lambda}_{\rm c}$ in several cases and 
show that it approaches zero
for scale-free directed networks under some additional assumptions. \\
\begin{center}
{\large \bf 4. Correlations between outdegrees and indegrees}  \\
\end{center}
\indent
\indent
Some of real complex networks contains a few hubs, that is, 
nodes with many outdegrees and indegrees, and
a vast nodes with very few degrees as well. 
In order to discuss such a phenomenon,
it is effective to introduce the conditional probability $p(\ell|k)$ for the
correlation between outdegrees and indegrees. 
It indicates the probability that a given $k$-indegree node has $\ell$ outdegrees. \\
\indent
First we deal with the following two extreme cases (I) and (II) by means of $p(\ell|k)$.
(I) has the highest correlation, while (II) is the lowest one or independent case. \\
\\
Case (I) ~~$p(\ell|k) = \delta_{\ell,k}$ for all $k$ and $\ell$.  \\
\\
\indent
Here $\delta_{\ell,k}$ is the Kronecker
delta. This condition implies that each node has the same in- and out-degrees.
If we regard $k$ and $\ell$ as random variables, then $k = \ell$.
In this case we have $\langle{k^2}\rangle = \langle{\ell^2}\rangle$ 
and the denominator in (8) is 
\[
\langle k{\ell}\rangle = \sum_{k, \ell} k\ell p(k, \ell) = \sum_{k, \ell} k\ell \delta_{\ell,k}P(k) = \langle{k^2}\rangle.
\]
Therefore, if the indegree distribution follows a power law $P(k) \propto k^{-\gamma}$
with $2 < \gamma \le 3$, then the threshold $\lambda_{\rm c}$ in (8) is equal to zero as in [1, 2, 6, 9],
where this prominent result for the SIR and SIS models was first obtained for undirected scale-free networks.
\\
\\
Case (II) ~~ $p(\ell|k) = Q(\ell)$ for all $k$ and $\ell$. \\
\\
\indent
In this case the random variables $k$ and $\ell$ are independent or uncorrelated and 
$\langle k\ell \rangle = \langle k \rangle \langle \ell \rangle$ holds,
from which we see that the threshold (8) is
\begin{eqnarray*}
    \lambda_{\rm c} = \frac {1}{{\langle{k}\rangle}} =  \frac {1}{{\langle{\ell}\rangle}}.
\end{eqnarray*}
This expression of the threshold also appears for a homogeneous 
SIS model in undirected networks as in [2]. \\
\indent
Under the conditions 
$\langle k^2 \rangle < \infty$ and $\langle {\ell}^2 \rangle < \infty$ 
the average $\langle k\ell \rangle$ satisfies
\begin{eqnarray*} 
\langle k\ell \rangle  \le \sqrt{\langle k^2 \rangle \langle {\ell}^2 \rangle},
\end{eqnarray*} 
by the Schwarz inequality [3]. Moreover, it also says 
that the equality $\langle k\ell \rangle  =
\sqrt{\langle k^2 \rangle \langle {\ell}^2 \rangle}$ holds only if both random variables $k$ and $\ell$
satisfy $k = a \ell$ with some constant $a$. From $\langle k \rangle = \langle \ell \rangle$ we see that
$a=1$, which coincides with (I). 
Thus, according as the correlation between indegrees $k$ and outdegrees $\ell$ becomes high,
the threshold $\lambda_{\rm c}$ approaches zero:
\[
   {\lambda}_{\rm c} = \frac{\langle{\ell}\rangle}{\langle{k\ell}\rangle} 
   \to \frac{\langle{\ell}\rangle}{\sqrt{\langle k^2 \rangle \langle {\ell}^2 \rangle}} \to 0,
\]
provided the power laws 
\begin{eqnarray} 
  P(k) \propto k^{-\gamma}~~{\rm and }~~ Q(\ell) \propto {\ell}^{-\gamma'}~~{\rm for } ~~ k, \ell \le M,
\end{eqnarray} 
hold with exponents $2 < \gamma, \gamma' \le 3$ and the maximum degree $M$ is very large. \\
\indent
In order to discuss more quantitatively, we might use the correlation coefficient ([3]):
\begin{eqnarray*} 
r = \frac{\langle (k-\langle k \rangle)(\ell -\langle \ell \rangle) \rangle}{\sigma_k \sigma_{\ell}}  
= \frac{\langle k \ell \rangle - \langle k \rangle \langle \ell \rangle}
{\sqrt{\langle k^2 \rangle- {\langle k \rangle}^2}
\sqrt{\langle {\ell}^2 \rangle- {\langle {\ell} \rangle}^2}}.
\end{eqnarray*}
Here $\sigma_k$ and $\sigma_{\ell}$ are the respective standard deviations. It satisfies 
$-1 \le r \le 1$, which is a variation of the Schwarz inequality.
By a simple calculation it follows that the threshold of (8) can be written as
\begin{eqnarray*} 
\lambda_{\rm c} = 
\Big(\langle k \rangle + r\langle k \rangle \sqrt{\big(\langle {k^2} \rangle /{\langle k \rangle}^2-1\big)
\big(\langle {\ell}^2 \rangle/{\langle {\ell} \rangle}^2-1\big)}\Big)^{-1}.
\end{eqnarray*} 
So, if it is possible to find $r > 0$ by sampling, then we have $\lambda_{\rm c} \approx 0$ under
the above condition (11), because $\langle {k^2} \rangle \gg {\langle k \rangle}^2$
and $\langle {{\ell}^2} \rangle \gg {\langle {\ell} \rangle}^2$ in case of $2 < \gamma, \gamma' \le 3$. \\
\begin{center}
{\bf\large References}
\end{center}
\begin{itemize}
\item[{[1]}]  
M. Barth\'elemy, A. Barrat, R. Pastor-Satorras and A. Vespignani,
Dynamical patterns of epidemic outbreaks in complex heterogeneous networks,
{\it Journal of Theoretical Biology} {\bf 235}, 275--288, 2005.
\vspace{-2mm}
\item[{[2]}] M. Bogu\~n\'a, R. Pastor-Satorras and A. Vespignani,
Epidemic spreading in complex networks with degree correlations,
LN in Physics {\bf 625}, Springer, 127--147, 2003.
\vspace{-2mm}
\item[{[3]}] Y. S. Chow and H. Teicher, Probability Theory, Springer Verlag, 1988.
\vspace{-2mm}
\item[{[4]}] S. N. Dorogovtsev and J. F. F. Mendes, Evolution
of Networks: From Biological Nets to the
Internet and WWW, Oxford University Press, 2003.
\vspace{-2mm}
\item[{[5]}] L. A. Meyers, M. E. J. Newman and B. Pourbohloul,
Predicting epidemics on directed contact networks, {\it Journal of Theoretical Biology} {\bf 240}, 400-418, 2006.
\vspace{-2mm}
\item[{[6]}] Y. Moreno, R. Pastor-Satorras and A. Vespignani, 
Epidemic outbreaks in complex heterogeneous networks, {\it European Physical Journal B} 
{\bf 26}, 521-529, 2002.
\vspace{-2mm}
\item[{[7]}] J. D. Murray, Mathematical Biology (Vol. 1), Springer Verlag, 2002.
\vspace{-2mm}
\item[{[8]}]  M. E. J. Newman, The structure and function of complex networks,
{\it SIAM Review} {\bf 45} 167--256, 2003.     
\vspace{-2mm}
\item[{[9]}]  R. Pastor-Satorras and A. Vespignani,
Epidemic spreading in scale-free networks, {\it Physical Review Letters} {\bf 86}, 3200--3203, 2001.
\vspace{-2mm}
\item[{[10]}]  S. Tanimoto, Power laws of the in-degree and out-degree distributions
of complex networks, arXiv:0912.2793, 2009.
\end{itemize}
\end{multicols}
\end{document}